\documentstyle[psfig,rotate]{mn}
\newcommand{\lsim}{{\, \lower2truept\hbox{
${< \atop\hbox{\raise4truept\hbox{$\sim$}}}$}\,}}
\newcommand{\gsim}{{\, \lower2truept\hbox{
${> \atop\hbox{\raise4truept\hbox{$\sim$}}}$}\,}}

\begin{document}

\title [Globular Cluster Calibration of SN 1992A]
{Globular Cluster Calibration of the Peak Brightness
of the Type Ia Supernova 1992A and the Value of H$_\circ$}

\author[M. Della Valle et al.]
{M. Della Valle$^1$, M. Kissler-Patig$^2$, J. Danziger$^3$ \&
J. Storm$^{4,5}$\\
$^1$Department of Astronomy, University of Padova, Vicolo dell'Osservatorio
5, 35122, Padova \\ 
$^2$UCO/Lick Observatory, University of California, Santa Cruz, CA 95064, USA\\
$^3$Osservatorio Astronomico, Trieste, via Tiepolo 11, 34131, Trieste\\
$^4$Astrophysikalisches Institut Potsdam, An der Sternwarte 16, D-14482, 
Potsdam, Germany\\
$^5$European Southern Observatory, Alonso de Cordoba 3107, Vitacura, Casilla 
19001, Santiago, Chile\\}
\maketitle

\begin{abstract} We have determined the absolute magnitude at maximum light of
SN 1992A by using the turn--over magnitude of the Globular Cluster
Luminosity Function of its parent galaxy, NGC 1380.  A recalibration
of the peak of the turn--over magnitude of the Milky Way clusters
using the latest HIPPARCOS results has been made with an assessment of
the complete random and systematic error budget.  The following
results emerge: a distance to NGC 1380 of 18.6$\pm 1.4$ Mpc,
corresponding to (m--M)=31.35$\pm 0.16$, and an absolute magnitude of
SN 1992A at maximum of M$_B{\rm (max)}=-18.79\pm0.16$.  Taken at face
value, SN 1992A seems to be more than half a magnitude fainter than
the other SNeI-a for which accurate distances exist. We discuss the
implications of this result and present possible explanations.  We
also discuss the Phillips's (1993) relationship between rate of
decline and the absolute magnitude at maximum, on the basis of 9
SNeI-a, whose individual distances have been obtained with Cepheids
and the Globular Cluster Luminosity Function.  The new calibration of
this relationship, applied to the most distant SNe of the Calan-Tololo
survey, yields H$_\circ= 62\pm 6$ km s$^{-1}$ Mpc$^{-1}$.
\end{abstract}


\section{Introduction}
\bigskip

The study of Supernovae Ia at maximum light is important for two reasons. 

On the one hand Supernovae Ia are commonly regarded as reliable
standard candles.  In the recent past, mainly on the basis of
photographic data, a number of authors [e.g. Leibundgut and Tammann
(1990), Miller and Branch (1990), Della Valle and Panagia (1992),
Vaughan et al. (1995)] were able to show that the absolute magnitudes
at maximum of SNeI-a have a small dispersion, of the order of $\lsim
0.3$ mag. In principle, used as standard candles, these objects could
provide distance measurements with an uncertainty of only $\lsim \pm
14\%$.  However, `high--quality' observations, obtained in the last
6--7 years, appear to complicate the previous `idyllic'
scenario. Phillips (1993), on the basis of a sample of SNeI-a whose
light-curves were well sampled at maximum light, has considerably
strengthened a former suggestion by Pskovskii (1967) concerning the
possible existence of a relationship between the Absolute Magnitude at
Maximum of type Ia Supernovae and their Rate of Decline
(=AMMRD). Branch, Romanishin and Baron (1996) have found that SNeI-a
occurring in early type galaxies are {\underline {on average}} $\sim
0.3$ magnitudes fainter than Ia in spirals.  Finally, Sandage et
al. (1996) were not able to confirm the Phillips's relationship. As a
consequence of these uncertainties, the value of {\sl H$_\circ$}
measured with SNeI-a varies between $\sim 50 $ and almost $70 $ km
s$^{-1}$ Mpc$^{-1}$ (Lanoix 1998, Hamuy et al. 1995).

On the other hand, a number of observational studies have pointed out
the existence of significant intrinsic differences between SNeI-a
occurring in spirals and early type galaxies: these observations
concern their spectroscopic (Branch, Drucker and Jeffery 1988,
Filippenko 1989, Branch and van den Bergh 1993, Nugent et al. 1994),
and photometric evolution (van den Bergh and Pierce 1992, van den
Bergh and Pazder 1992, Suntzeff 1996, Riess et al. 1996), their rate
and place of occurrence (e.g. Bartunov et al. 1994, Della Valle and
Livio 1994, Cappellaro et al. 1997a, Wang et al. 1997), all of which
question the uniqueness of the progenitors for type Ia Supernovae.
Since the peak luminosity is proportional to the synthesized nickel
mass (e.g. Cappellaro et al. 1997b), one of the most direct
observational ways to prove the existence of intrinsic differences
between the progenitors is to measure the differences in absolute
magnitude at maximum for a sample of SNeI-a.  Such differences might
then be attributed to either differences among the progenitors (Branch
et al. 1995 for a review) and/or to differences in the mechanism of
the explosion (e.g. Canal, Isern and Lopez 1988).
 
These problems have motivated our study of SN 1992A. In section
\ref{sec.gclf}, we will briefly discuss the data and methods which
we have used to determine the turn--over magnitude of the Globular
Cluster Luminosity Function (=GCLF) of NGC 1380.  In section
\ref{sec.distance}, we determine the distance to NGC 1380 and in
section \ref{sec.abs_mag} we derive the absolute magnitude at maximum
light for SN 1992A.  In section
\ref{sec.abs_mag_general} we determine the AMMRD. In section
\ref{sec.discussion} we discuss the results and the possible
implications for the calibration of the extragalactic distance scale
and in the final section (\ref{sec.conclusion}) we summarize our
conclusions.
\bigskip

\section{The Globular Cluster Luminosity Function}
\label{sec.gclf}
\bigskip

\subsection{The data}
\label{subsec.data}

  The GCLF was determined on the basis of data obtained at the ESO New
  Technology Telescope (NTT) using the ESO Multi-Mode Instrument
  (EMMI) in December 1993.  The stacked frames of NGC 1380 totaled an
  equivalent exposure time of 2700 sec in $R$, 7800 sec in $V$, and
  15000 sec in $B$ with an effective FWHM of stellar images $\lsim
  1.0$ arcsec in all filters. The corrected field of view covered by
  all three filters was approximately 6.5 by 6.5 arcminutes.

A comprehensive description of the data reduction is given in
Kissler-Patig et al. (1997), where we fully analyze the data on the
globular cluster systems. In the present paper we concentrate on the
turn-over magnitude of the globular cluster luminosity functions.

Briefly, we modeled
the galaxy light with isophotal models in all colours and subtracted it
from the original frames (Fig. 1). 
The photometry was carried out on these flat
background exposures with DAOPHOT operating in IRAF. Artificial star
experiments with typically 20000 stars added in each colour over many
runs were carried out to determine our finding completeness.  Finally
our $B$, $V$,and $R$ samples were combined (roughly 900 objects), point
sources were selected (710 objects) and then selected in colour (422
objects) to reject fore- and back-ground objects.  The
globular cluster luminosity function was then computed with objects
further than 35 arcsec and closer than 125 arcsec to the center of the
galaxy. The lower limit avoids incompleteness changes towards the
center, the upper limit was the limit of the surface density profile
of the globular clusters.
Finally we linked our photometry for the globular clusters to the one
used to derive the lightcurve of SN 1992A by adjusting our calibration 
to match the comparison stars used by Cappellaro et al. (1997b). This required 
systematic shifts of +0.070 to +0.095 in all colours after which the RMS 
scatter for the comparison stars was 0.02~mag.

\subsection{Computing the luminosity function}
\label{subsec.comp_gclf}

We binned the globular clusters into 0.35 mag bins, and plotted their
number versus magnitude to obtain the luminosity function.  These
luminosity functions in $B$, $V$, and $R$ are shown in Figs. 2, 3,
and 4. The counts are given in Tab. 1, together with the completeness
that we corrected for. Note that no completeness correction exceeding
10\% was necessary until beyond the turn--over. Furthermore we only
considered globular clusters down to our 60\% completeness limit (25.3
mag in $B$, 24.2 mag in $V$, 23.9 mag in $R$) in the fits.
\begin{table}
\caption{Globular cluster luminosity functions in $B$, $V$, and $R$
with the completeness factor} 
\begin{tabular}{lllllllll}
\hline
$B$ & counts &  comp  & $V$ & counts &  comp & $R$ & counts
&  comp \\
\hline
   20.35 &   0. & 1. &  20.35 &   0. &     1. & 20.35&        0. &         1.\\
    20.7 &   0. & 1. &   20.7 &   0. &     1. &  20.7&        2. &         1.\\
   21.05 &   0. & 1. &  21.05 &   1. &     1. & 21.05&        7. &         1.\\
    21.4 &   0. & 1. &   21.4 &   5. &     1. &  21.4&       8. &         1.\\
   21.75 &   1. & 1. &  21.75 &   7. &     1. & 21.75&  30. &   1.\\
    22.1 &   2. &   1. &   22.1 &  17. &     1. &  22.1&  35.4 &  0.96\\
   22.45 &   7. &   1. &  22.45 & 28.0 &   0.96 & 22.45&  46.0 &  0.96\\
    22.8 &  15. &   1. &   22.8 & 45.0 &  0.96 &  22.8&  48.3 &   0.93\\
   23.15 &  26.9 &  0.97 &  23.15 & 56.3 &   0.94 & 23.15&  80.4 &  0.91\\
    23.5 &   40.8 &  0.96 &   23.5 & 56.4 &  0.92 &  23.5&  46.3 &  0.84\\
   23.85 & 47.3 &  0.95 &  23.85 & 66.6 &  0.87 & 23.85&  45.3 &  0.70\\
    24.2 & 73.4 &   0.93 &   24.2 & 52.5 &  0.78 &  24.2&  46.6 &  0.45\\
   24.55 & 52.8 &  0.89 &  24.55 & 48.6 &  0.60 & 24.55&  17.8 &  0.17\\
    24.9 & 59.0 &  0.81 &   24.9 & 23.0 &  0.26 &  24.9&   --   &  -- \\
   25.25 & 39.4 &  0.66 &  25.25 &  --   &  -- & 25.25&   --   &  -- \\
    25.6 & 42.2 &  0.35 &   25.6 &   -- &  -- &  25.6& -- &  -- \\
\hline
\end{tabular}
\end{table}

\subsection{Turn--over values in $B$, $V$, and $R$}

The globular cluster luminosity function is well 
represented by a Gaussian (Whitmore 1997 for the most recent review), or
a $t_5$ function (Secker 1992, Kissler et al.~1994), with a universal absolute
peak luminosity (see next paragraph). We fitted both types
of functions to our luminosity functions to derive the turn--over
magnitudes and width simultaneously. 

The robustness of our result are tested in several ways.  
We experimented with different binnings, varying the bin width and
bin centers. We used subsamples of our globular clusters and computed their
luminosity function. We fitted the luminosity function down to various
magnitudes. Finally we used different ``combined'' completeness
factors (i.e.~the
product of the completeness factors of different colours) as well as the
completeness for individual colours. In all cases we obtained similar
results for the turn--over magnitudes and distribution width 
within about 0.10 mag, 
independent of the type of function used. This gives us confidence
in the results and provides a measure of the overall error of less than
0.10 mag. The results are tabulated in Tab. 2.

\begin{table}
\label{tab.turnover}
\caption{Turn--over values and width for the globular cluster luminosity
function in $B$, $V$, and $R$} 
\begin{tabular}{lllll}
\hline
& \multicolumn{2}{c}{Gaussian} & \multicolumn{2}{c}{$t_5$ function}\\
&  T--O mag & $\sigma$ & T--O mag & $\sigma$ \\
\hline
$B$ &  $24.38\pm0.09$ & $0.92\pm0.10$ & $24.38\pm0.09$ & $0.89\pm0.10$ \\
$V$ &  $23.67\pm0.11$ & $0.96\pm0.10$ & $23.69\pm0.11$ & $0.95\pm0.10$ \\
$R$ &  $23.16\pm0.09$ & $1.00\pm0.10$ & $23.17\pm0.10$ & $0.98\pm0.10$ \\
\hline
\end{tabular}
\end{table}

Forcing a broad function ($\sigma_{Gauss}=1.35$) to our data,
as proposed by Whitmore (1997) for bright ellipticals, pushes the
turn--over values to $\approx 0.15$ magnitudes fainter but results in 
unaccetable fits. The width of the GCLF of NGC 1380 (an S0 galaxy) is
in much better agreement with values found for the Milky Way and M31 
(e.g.~Secker 1992) and fainter early--type galaxies (Kissler et al.~1994,
Kissler-Patig, Richtler, \& Hilker 1996), supporting the result from our free 
fits. 

A previous determination of the $V$ turn--over magnitude for NGC 1380 was made 
by Blakeslee \& Tonry (1996), who derived a value of $m_V^0=24.05\pm0.25$
in disagreement with our results. However, it seems (Blakeslee
priv.~com.~) that from their $V$ data alone, they could not correct for
a background cluster located about 30 arcsec East of NGC 1380 (see
Kissler-Patig et al.~1997). If we include these background galaxies
(having $V$ magnitudes between 21 and 24 mag), and compute the fit to their
magnitude limit, we can reproduce their
result in the sense that our peak appears to be shifted by almost half a 
magnitude fainter.  

\section{The distance to NGC 1380}
\label{sec.distance}
\bigskip

It is now well established that, excluding some small metallicity and
perhaps some galaxy--type effects, the absolute turn--over of the GCLF
is constant for the old clusters in all galaxies (e.g. Whitmore 1997, Harris 
1997).  
The absolute peak luminosity can be
derived from ``local'' calibrators such as the globular cluster
systems of the Milky Way or M31.  While comparing globular clusters in
the Milky Way and M31 to those in bright ellipticals is often quoted
as comparing ``apples with oranges'', because of possible
globular cluster population differences (``galaxy--type'' effects), the 
comparison makes more sense in the case of NGC 1380 classified as an S0 galaxy.
The colour distribution of the NGC 1380 globular clusters is similar to
that of the Milky Way globular clusters (Kissler-Patig et al.~1997),
including old blue halo clusters, and old red "bulge" clusters, with
only the colour distribution in NGC 1380 extending further to the red.
We shall shortly discuss the implications, but shall 
first review the absolute turn--over luminosities for the
Milky Way and M31 GCLF's.

Secker (1992) and Sandage \& Tammann (1995) recently determined
the turn--over values $M_V^0$ and $M_B^0$ for M31 and the Milky Way.
They respectively found $M_V^0=-7.51\pm0.15$ and $M_V^0=-7.29\pm0.13$, and
 $M_V^0=-7.70\pm0.20$ and $M_V^0=-7.60\pm0.11$ for M31 and the Milky Way.
While the peak of the GCLF in M31 depends ``only'' on the adopted distance for
Andromeda, the peak of the GCLF in the Milky Way depends on the 
distance moduli adopted for the individual clusters, i.e.~for our purpose
on the absolute magnitude of the horizontal branch and its metallicity
dependence. The former large uncertainties are reflected in the discrepant
values of Secker and Sandage \& Tammann. However, recently Reid (1997)
and Gratton et al.~(1997) derived distances to several globular
clusters by attaching their main sequences to nearby subdwarfs whose
parallaxes had been measured with high precision by {\sl HIPPARCOS}.
Their results support a dependence of the absolute horizontal branch
luminosity on the metallicity of the type:
$M_V(HB)=0.29(\pm0.09)\cdot({\rm [Fe/H]} + 1.5) + 0.43(\pm 0.04)$. 
We therefore decided to re-fit the
GCLF of the Milky Way in $B$, $V$, and $R$ using the new distance implied
by this relation.

We used the McMaster globular cluster database (Harris 1996)
compiling data for 146 globular clusters in the Milky Way. We corrected for
reddening according to Rieke \& Lebofsky (1985), and derived distances
using the relation mentioned above to derive absolute $B$, $V$ and $R$
magnitudes for all clusters.
The resulting luminosity functions are shown in Fig.~5,
together with the best Gaussian fits. The mean from
various fits to different binnings with both Gaussian and $t_5$ functions,
and various subsamples (excluding the most reddened clusters),  
are listed in Tab.~\ref{tab.mw_turnover}, the errors are given as
dispersions around the mean, without considering any systematic error. 
The errors in the individual determinations are typically of the order 0.07
mag, as shown in the table. Varying the 
binning and subsamples results in typical
errors of the same order, leading to a total random error of 0.10 mag.
However we have to account for any systematic error introduced by our
$M_V(HB)$--[Fe/H] relation. We therefore recomputed all distances for 
Fusi--Pecci's et al.~(1996) relation, derived however from M31 clusters,
and implying a much weaker metallicity dependence [$M_V(HB)=0.13(\pm0.07)
\cdot({\rm [Fe/H]}) + 0.95(\pm 0.09)$] (see also Carney et al. 1992).
We then derive mean turn--over
values systematically {\it fainter} by about 0.2 mag. The above two relations
span the range currently under debate. In the following we will adopt the
relation derived for the Milky Way clusters using the Gratton et al. (1997)
relation, and keep in mind a possible
systematic error of the turn--over values of up to --0.2 mag.   
\begin{table}
\label{tab.mw_turnover}
\caption{Turn--over values and dispersions of the Milky Way globular
cluster luminosity function in $B$, $V$, and $R$ derived from the
McMaster University globular cluster database}
\begin{tabular}{l l l}
\hline
Colour & TO value & $\sigma_{Gauss}$ \\
\hline
$B$ & $-7.08\pm0.08$ & $1.07\pm0.08$ \\
$V$ & $-7.62\pm0.06$ & $1.16\pm0.09$ \\
$R$ & $-8.14\pm0.07$ & $0.88\pm0.11$ \\
\hline
\end{tabular}
\end{table}

Our final result for the Milky Way [$M_V^0=-7.62\pm0.10$] compares
well with the average of the
values derived for M31 by Secker (1992) and Sandage \& Tammann (1995)
($M_V^0=-7.61\pm0.13$).
By combining our Milky Way results with the turn--over magnitudes of 
Tab. 2, we find $(m-M)_V=31.29\pm 0.14$, $(m-M)_B=31.46\pm 
0.13$, and $(m-M)_R=31.30\pm 0.13$, (errors being random errors, not
including up to --0.2 mag systematic error).
 
Two related effects have to be further considered: the metallicity effect
on the turn--over (metal rich globular cluster having fainter absolute
magnitude), and a galaxy--type effect. While the first one was well
quantified by Ashman, Conti \& Zepf (1995), the second effect rests on a
less firm base (Harris 1997), since it implies known distances to galaxies
of different types with well observed GCLFs. While central giant ellipticals 
might differ from other galaxies due to their complex globular cluster
system, it is not clear at all that GCLFs in spirals and small ellipticals
differ systematically. The effect would be of the order of 0.1 mag between
an S0 and a spiral in the sense that the S0 would have a fainter turn--over
magnitude. We shall however not correct for this effect but include it in the 
systematic error budget. 

The metallicity dependence of the turn--over is better quantified (Ashman, 
Conti \& Zepf 1995). In the Milky Way, the GCLF is dominated by clusters
with [Fe/H]$\simeq -1.4$ dex. In NGC 1380, Kissler-Patig et al.~(1997)
identified two populations of globular clusters. One matching the halo
population of the Milky Way, the other being more metal rich
([Fe/H]=$-0.2\pm0.5$). As we used one sample (mean [Fe/H]=$-0.6\pm0.5$)
including both populations to derive the GCLF above, based on metallicity 
alone, the correction to apply to $M_V^0$ should be of the order of 
$0.2\pm 0.1$ to fainter magnitudes. However, as shown in Kissler-Patig et
al.~(1997) the two populations have identical turn--overs in $V$,
compatible (also in $B$ and $R$) with a slightly younger red population, 
compensating the metallicity effect with an age effect. 
The correction should therefore be reduced to less than 0.1 mag, that we
neglect and include it in our random errors.

In summary, we end up with a mean distance modulus of $(m-M)=31.35\pm0.16$,
keeping in mind that two possible systematic errors 
(different $M_V(HB)$--[Fe/H]
relation, weak galaxy type effect between S0 and spirals) 
are not included, but could only act to {\it lower} the distance 
modulus up to 0.3 mag. The distance modulus corresponds to a distance of 
$18.6\pm 1.4$~Mpc. 

Further comparison can be made with the turn--over values and
distances (using the above calibration) of other Fornax members 
(Tab. 4).

\begin{table*}
\label{tab.gal_dist}
\caption{Distance to individual galaxies in the Fornax cluster}
\begin{tabular}{lllcll}
\hline
Galaxy & TO(V) & $(m-M)_V$ & Hubble type & Method & References\\
\hline
NGC 1344 & $23.80\pm0.25$ & $31.42\pm0.32$ & E & GCLF & 1 \\
NGC 1365 & --             & $31.32\pm0.20$ & SBb & Cepheids& 2 \\ 
NGC 1374 & $23.48\pm0.15$ & $31.10\pm0.25$ & E & GCLF & 3 \\
NGC 1379 & $23.66\pm0.30$ & $31.28\pm0.37$ & E & GCLF & 3\\
NGC 1380 & $23.67\pm0.13$ & $31.29\pm0.16$ & S0& GCLF & this work \\
NGC 1399 & $23.81\pm0.09$ & $31.43\pm0.22$ & E (Cd?) & GCLF & 4\\
NGC 1404 & $24.01\pm0.14$ & $31.63\pm0.24$ & E & GCLF & 1,5 \\
NGC 1427 & $23.70\pm0.20$ & $31.32\pm0.28$ & E & GCLF & 3 \\
\hline
\end{tabular}

{\scriptsize 1: Blakeslee \& Tonry, 2: Madore et al. 1996, 3: Kohle et al. 
1996, 4: Whitmore 1997, 5: Richtler et al. 1992. }
\end{table*}

The weighted mean of the distances of Tab. 4
gives $(m-M)=31.35\pm 0.09$ ($1\sigma)$ or $D=18.6\pm 0.8$ Mpc.
The comparison between the $\pm 3\sigma$ range with each distance, 
might suggest that these members form a distribution 
of Fornax cluster elongated 
along the line of sight by less than $\pm 14\%$, but all the derived
distance moduli include the mean within the errors supporting previous
claims that Fornax is a very compact cluster.

The $V$ turn--over magnitude for NGC 1380 falls well within
the range spanned by the other Fornax galaxies. Excluding NGC 1399, the
central giant cD for the reasons given above (e.g.~Harris 1997), and the
second brightest galaxy NGC 1404 (see Richtler et al.~1992 for its peculiar
situation), we note that the turn--over magnitude of the smaller galaxies is
strikingly constant, supporting a weak dependence of the turn--over
luminosity with type among these galaxies.

We note in passing that our derived distance modulus of $(m-M)=31.35\pm0.16$ 
falls precisly at the mean
of Tab. \ref{tab.gal_dist}, and agrees perfectly with the Cepheid distance
to NGC 1365. 
\bigskip

\section{The absolute magnitude at maximum of SN 1992A}
\label{sec.abs_mag}

SN 1992A is one of a few type Ia supernovae which have been discovered before 
maximum light, therefore
the apparent magnitude at maximum light has been measured with great accuracy.
The $B$ maximum occurred on January 19, 1992 at $B=12.56\pm0.02$ and
$(B-V)=0.00\pm 0.02$ (Suntzeff 1996). By assuming 
$(m-M)=31.35\pm 0.16$, we find an absolute magnitude at
maximum of $M_B=-18.79\pm0.16$. This result suggests different  
possibilities.

Branch, Romanishin and Baron (1996) have recently established a number
of statistical correlations between the parameters derived from the
analysis of the light-curves and spectral evolution of SNeI-a
[e.g. B(max)--V(max), rate of decline, expansion velocity of SiII
absorption line] and the colour of the respective parent
galaxies. These authors find that the SNeI-a in early type galaxies
are on average (even after disregarding sub-luminous objects such as
1991bg) fainter than Ia in `spirals' by $\sim 0.3\pm0.1$ with
$\sigma_{M_B(max)}=0.23$.  By taking, as zero point of the
calibration, the absolute magnitude of SNeI-a in late type galaxies,
i.e. $M_B=-19.53\pm0.07$ (Tammann et al. 1996), obtained via Cepheid
distances with HST, it seems that Ia in `early' type galaxies should
have $M_B \approx -19.2\pm0.2$. With these figures taken at face value,
SN 1992A might be about one half magnitude fainter than expected, and therefore
one could conjecture about its possible peculiarity. However we note
the following. Subluminous SNe such as SN 1991bg (Filippenko et al. 1992a,
Leibundgut et al. 1993, Turatto et al. 1996 )
or SN 1992K (Hamuy et al. 1994) are clearly identifiable as peculiar
objects from their own spectroscopic evolution (e.g. redder continuum,
lower photosperic expansion velocity).  This is not the case for SN
1992A.  Indeed its spectroscopic evolution (Kirshner et al 1993,
Suntzeff 1996), is typical of {\sl normal} SNeI-a (e.g. Branch,
Fisher, Nugent 1993).  In addition, Patat et al. (1996) have shown, in
a convincing way (see their Fig. 8 ), that the spectroscopic evolution 
(at early stages) of SN 1992A is almost identical to that of the 
spectroscopically
normal SN 1994D. The photometric data provide a quite
contradictory picture. SN 1994D achieved at maximum B=11.60 and V=11.72
after the small
correction to the apparent magnitude at maximum to account for an
E(B--V)=0.06. Unfortunately
the distance to its early type (E7) parent galaxy, NGC 4526, has not
been measured directly. One might tentatively assume that the distance
of NGC 4526, a {\sl bona fide} member of the Virgo cluster (Sandage,
Binggeli et Tammann 1985), presumably ranges between 17.1 Mpc
of NGC 4321 (Freedman et al.  1994) and 25.5 Mpc of 4639
(Sandage et al. 1996). 
Both measures are based on the P-L relationship
of Cepheids.  After applying these distance moduli (with the attached
errors) to the B mag at maximum of SN 1994D, one concludes that the
absolute magnitude at maximum of SN 1994D presumably falls between
M$_B=-19.34$ and M$_B=-20.65$.  Although this range of
magnitude is quite large, this result would indicate that the
magnitude at maximum of SN 1994D is consistent with either the average
absolute at maximum derived above for SNeI-a in early-type galaxies or
with the possibility that this SN could have been unusually bright at
maximum.  In all cases (assuming that NGC 4526 is a member of the Virgo
cluster) the peak of light of SN 1994D is {\underline {definitely
brighter}} than that exhibited by SN 1992A. This implies that SNeI-a
which appear almost {\sl spectroscopically identical} (at early stages), 
can still span
a range of maximum brightness of $\gsim 0.5$~mag.
\bigskip

\section{The absolute magnitude at maximum of SNeI-a in early type
galaxies}
\label{sec.abs_mag_general}

Differences in the photometric evolution of type I SNe were first
pointed out by Barbon et al. (1973) and Pskovskii
(1967). The latter author also suggested that the absolute
magnitude at maximum of type I SNe correlates with the rate of decline
($\beta$) in such a way that: the faster the early decline of a SN
the fainter its absolute magnitude at maximum.  However, owing to the
difficulties in measuring $\beta$ even in the best observed objects
(e.g. Hamuy et al. 1991), such a relationship has been the subject of
debate for some time. Phillips (1993), by using as a tracer of different
rates of decline the total drop in magnitude that a SN undergoes
from its peak brightness until 15 days after maximum light
[$\Delta m_{15}(B)$], has considerably improved the situation (see
also Maza et al. 1994 and Hamuy et al. 1995).  From his Fig. 1 (top) a
clear relationship between $\Delta m_{15}(B)$ and the absolute magnitude
at maximum seems to exist. However, Sandage et al. (1996), by studying
the absolute magnitude at maximum and the rates of decline of a sample
of type Ia SNe occurring in late-type galaxies concluded
that {\sl There is no clear relation between the absolute
magnitudes.... and the decay rates} (see also H\"oflich et al. 1996).
The existence of this discrepancy
motivates our analysis.  We have collected from the literature (Tab. 5) the 
following data for SNeI-a fulfilling the following requirements: 
a) to have been
observed at maximum (or near maximum) with a reasonable
photometric error; b) the {\underline {individual}} distances to the
parent galaxies to be measurable via Cepheids (for spirals) and GCLF
(for early types). In addition we have determined (from the original
data) the apparent magnitude and the rates of decline of 3 `historical'
SNe (SN 1919A, 1939A and 1957B) because the TO magnitudes of the GCLF
of their parent galaxies 
have been determined, in recent years (see Whitmore 1997 and references
therein).

The selected objects are listed in Tab. 5 which gives the SN
designation with the parent galaxy (col. 1); the apparent magnitude at
maximum (col. 2); the rate of decline (col. 3); the adopted turn-over
magnitude (col. 4); the distance modulus derived by using our TO value,
see Tab. 3, (col. 5); the absolute B magnitude\footnote{The 
photographic magnitudes have been reduced to the B band through the colour
equations provided by Arp (1956)}  at maximum (col. 6) 
and the references (col. 7).

For SNe Ia calibrated by Cepheids, i.e. 1937C (Sandage et al. 1992,
Saha et al. 1994),
SN 1972E (Saha et al. 1995), SN 1981B (Saha et al. 1996a), SN 1960F
(Saha et al. 1996b), SN 1990N (Leibundgut et al. 1991, 
Sandage et al 1996), we have extracted
the B absolute magnitude at maximum from Tammann (1996) [for 1981B we
have used the re-calibrated apparent magnitude at maximum (Patat et al.
1996), for SN 1990N the new data come from Lira et al. (1998)], 
the $\Delta m(15)$ are from van den Bergh (1996) and Schaefer
(1996a, 1996b, 1996c).

\begin{table*}  
\caption{Data for SNeI-a whose distance has been measured
through the use of the peak of the 
GCLF}
\small {

\begin{tabular}{lllllll}
\hline
SN & m(max) & $\Delta m(15)$ & TO(V) & $(m-M)_\circ$ & M$_B$ & Ref. \\
\hline
1919A({\small {N 4486}}) & 12.2(pg)$\pm 0.2$ &  1.15$\pm 0.10$ & $23.88\pm 0.07$ & 31.50$\pm
0.21$&$-19.1\pm 0.3$   & 1,2,3,4,16\\
1939A({\small {N 4636}}) & 12.4(pg)$\pm 0.35$ & 1.5$\pm 0.2$ & $24.18\pm 0.20$ & 31.77$\pm 0.28$
& $-19.2\pm 0.45$ & 5,6,7,8,16 \\
1957B({\small {N 4374}}) & 12.4(pg)$\pm0.3$ & 1.3$\pm 0.2$ & $24.03\pm 0.3$  & $31.65\pm 0.36$
&$-19.05\pm 0.47$ & 9,10,11,12,13,16\\
1980I({\small {N 4374}})& 12.7B$\pm0.1$ & -- & $24.03\pm 0.3$  & $31.65\pm 0.36$
&$-18.95\pm 0.37$ & 16,17\\
1991bg({\small {N 4374}}) & 14.75(B)$\pm 0.1$ & 1.95$\pm 0.05$ & $24.03\pm 0.3 $& $31.65\pm 0.36$
&$-16.90\pm 0.37$ & 13,14,16\\
1992A({\small {N 1380}}) & 12.56(B)$\pm 0.02$ & 1.47$\pm 0.05$ &$ 23.67\pm 0.11$  & $31.35\pm 0.16$ &
$-18.79\pm 0.16$ & 15,16\\
unweighted mean$^\star$ & & & & & $-19.02\pm 0.16$ & \\
weighted mean$^\star$ & & & & & $-18.91\pm 0.12$ & \\
\hline
\end{tabular}
}

{\scriptsize 1: Baade 1936, 2: Baade 1938, 1941, 3: van den Bergh 1961, 
4: Whitemore 1997, 5: Baade 1941, 6: Hoffleit 1939, 7: Giclas 1939, 8:
Kissler et al. 1994, 9: Bertola 1964, 10: G\"otz 1958, 11: Romano
1957, 12: Li Tzin 1957, 13: distance derived from Ajhar et al.
1994, 14: Turatto et al. 1996, 15: Suntzeff 1996, 16: this paper, 
17: Barbon et al. 1989. \\
$^\star$ after excluding SN 1991bg}
\end{table*}

Fig. 6 shows the plot of the absolute magnitude at maximum M$_B$ as a
function of $\Delta m(15)$. Circles represent SNe in Spirals,
triangles SNe in early type galaxies.  We have computed the best-fit
regression by using a least squares fit, taking
into account the errors in both coordinates, and we obtain:
${\rm M_B}=2.40\pm 0.33(1\sigma) \times \Delta m(15)-22.06\pm0.41(1\sigma)$

Light symbols in Fig. 6 (not used in computing the fit) represent
SNe of Tab.\ref{tab.fornax_SNe} for which no individual distances to
the parent galaxies, obtained via GCLF or Cepheids, are available.

The slope of our fit is similar to that obtained by Phillips (1993),
though only 4 objects are in common (1990N, 1981B, 1991bg and 1992A). The
Phillips's intercept is fainter by $\sim $ 0.4 magnitudes, possibly
reflecting differences in the zero point of the methods used to
determine the distances to each parent galaxy. i.e. Surface Brightness
Fluctuation and Tully-Fisher in Phillips and Cepheids and GCLF
in this paper. 

If one restricts the analysis only to the prototypical SNeI-a [ as
defined by Branch, Fisher and Nugent (1993) or Branch and Tammann
(1992)] which are the only ones actually usable to determine the
distances, one has to perform the fit after excluding SN 1991bg 
(dashed line in Fig. 6). In this case we obtain a
considerably less steep slope and a fainter intercept 
${\rm M_B}=1.52\pm 0.42(1\sigma) \times \Delta m(15)-21.07\pm0.49(1\sigma)$.
However, these data can be more profitably used 
to improve the zero point of the relationship
found by Hamuy et al. (1996) whose slope has been determined on 26 SNe
whereas the zero point is based on 4 SNeI-a (1937C, 1972E, 1981B, 1990N)
which all appeared only in late type galaxies. 
We have fitted our data points by adopting the Hamuy's slope (0.784) 
and assuming as zero point the value of the intercept which minimizes the 
reduced $\chi^2$ (1.03). This fit (solid line in Fig. 6) yields: 
$${\rm M_B}=0.784\pm 0.18(1\sigma) \times \Delta m(15)-20.24\pm 0.24(1\sigma)~~~~~~~[1]$$
\bigskip

\section{Discussion}
\label{sec.discussion}
\bigskip

Two other SNeI-a have appeared in the Fornax cluster: 1980N and 1981D,
both in NGC 1316. Their peak luminosities were B(max)=12.49 and B(max)=12.59
(Hamuy et al. 1991, Sandage and Tammann 1996). 

\begin{table}
\label{tab.fornax_SNe} 
\caption{ Data for SNe for which no distance to the parent galaxy via Cepheids or GCLF
is known.} 
\begin{tabular}{llll}
\hline
SN & M$_B$ & $\Delta m(15)$ & Ref. \\
\hline
1980N &  --18.86$\pm 0.29$ &  1.20$\pm 0.10$ &1,2\\
1981D &  --18.76$\pm 0.29$ &  1.15$\pm 0.10$ &1,2\\
1989B &  --19.51$\pm 0.21$ &  1.31$\pm 0.07$ &3,4\\
1991T &  --19.94$\pm 0.23$ &  0.94$\pm 0.07$ &5,6\\
1994D &  --19.88$\pm 0.66$ &  1.26$\pm 0.05$ &7\\
\hline
\end{tabular}

{\scriptsize 1: Sandage \& Tammann 1996, 2: Hamuy et al. 1991, 
3: Tammann et al. 1996, 4: Wells et al. 1994, 5: van den Bergh 1996, 
6: Phillips 1993, 7: Patat et al. 1996}
\end{table}

The only direct measurement of the distance of their parent galaxy comes
from the Planetary Nebula Luminosity Function (=PNLF; see Tab. 7).  By
combining the distance modulus provided by McMillan, Ciardullo and
Jacoby (1993) and Jacoby (1997) with the respective apparent
magnitudes at maximum, we derive M$_B=-18.69\pm0.12$ and
M$_B=-18.59\pm0.12$, in good agreement with M$_B=-18.79\pm0.16$ of SN
1992A.

This result suggests a number of possible alternative interpretations: 

\begin{enumerate}

\item Spectroscopic normal Supernovae Ia in early type galaxies 
can be fainter, by $\sim 0.7-0.8$ mag, than Supernovae in Spirals
(see also van den Bergh and Pazder 1992).

\item As an alternative one might consider that the absolute magnitudes at maximum
of SNeI-a adopted in this paper for late type galaxies are `too
bright'. Since the Sandage et al. group makes use of multicolor photometry
to determine the P-L relationship of Cepheids, one can immediately
rule out the effect of reddening. On the other hand, the effect of the
metallicity on the P-L relation is still poorly known (e.g. Kennicutt et al. 1998).
Chiosi, Wood and Capitanio (1993) predict a small sensitivity of the P-L relationship
to metallicity, while a recent work by Beaulieu et al. (1997) might
even indicate an opposite trend (see also Efremov 1996).

\item The similarity of the absolute magnitude at maximum of 1992A 
with that of 1981D and 1980N might be only illusory. In fact, if one
compares the GCLF and PNLF distances of NGC 1399 and N 1404, both belonging to
Fornax, a difference, (m--M)$_{GC}$--(m--M)$_{PNe}\sim 0.3-0.4$ mag, 
between the distance scale zero points seems to exist (if we extend this
analysis to NGC 4374 and NGC 4486, belonging to the Virgo Cluster,
$\Delta m$ increases up to 0.6--0.7 mag). In turn this would imply that 
1981D and
1980N have M$_B ({\rm max})\sim -19$ and NGC 1316 should be placed
on the far-side of the Fornax cluster (like NGC 1404).
This possibility is not entirely ruled out considering that NGC 1316 is
located in the southern extension of the Fornax cluster and not in
the tight main concentration where all the other Fornax galaxies presented
in Tab.\ref{tab.gal_dist} are located. In this scenario only
1992A would be moderately sub-luminous.

\end{enumerate}

\begin{table*}
\label{tab.pne_dist}
\caption{Data for SNeI-a whose distance has been measured through
the use of the PNLF}
\begin{tabular}{lllccl}
\hline
SN   & Galaxy   & $(m-M)_\circ$   & M$_B$ & $\Delta m(15)$& References\\
\hline
1980N& NGC 1316 & $31.19\pm 0.07$ & $-18.69\pm 0.12$ & 1.2$\pm 0.1$  & 1,2,4,5 \\
1981D& NGC 1316 & $31.19\pm 0.07$ & $-18.59\pm 0.12$ & 1.15$\pm 0.1$ & 1,2,4,5 \\ 
1957B& NGC 4374 & $31.04\pm 0.18$ & $-18.53\pm 0.21$ & 1.3$\pm 0.3$  & 3,2,4,5 \\
1980I& NGC 4374 & $31.04\pm 0.18$ & $-18.33\pm 0.21$ &  --           & 3,2,4,5 \\
1991bg&NGC 4374 & $31.04\pm 0.18$ & $-16.13\pm 0.21$ & 1.95$\pm 0.05$& 3,2,4,5 \\ 
1960R& NGC 4382 & $30.85\pm 0.17$ & $-18.95\pm 0.25$ & 1.2$\pm 0.1 $ & 3,2,4,5,6\\
1919A& NGC 4486 & $30.79\pm 0.17$ & $-18.49\pm 0.21$ & 1.15$\pm 0.1$ & 2,4,5 \\
unweigthed mean$^\star$ &                 &                  & $-18.60\pm 0.21$ & & \\
weigthed mean$^\star$ &                 &                  & $-18.60\pm 0.07$ & & \\
\hline
\end{tabular}

{\scriptsize 1: McMillan, Ciardullo \& Jacoby (1993), 2: Jacoby (1997)
3: Jacoby, Ciardullo \& Ford (1990), 4: this paper, 5: Barbon et al. (1989), 
6: Bertola (1964), $^\star$after excluding SN 1991bg}

\end{table*}
\bigskip

\section{Conclusions}
\label{sec.conclusion}

In this paper we have determined the distance to NGC 1380, an S0 galaxy
host of the type Ia SN 1992A, through the use of the TO magnitude of
the GCLF. We find a distance modulus of 31.35$\pm 0.16$ corresponding
to a distance of 18.6$\pm 1.4$ Mpc.  This is consistent with the
distances to other members of the Fornax cluster, utilizing the same
method, as well as the Cepheid distance to NGC 1365.

By applying this distance to the apparent magnitude of SN
1992A, we find that at peak brightness SN 1992A reached $M_B=
-18.79\pm0.16$, which is about 0.4~mag fainter than expected
for typical SNeI-a in early type galaxies (Branch, Romanishin and
Baron 1996), and about 0.7 magnitudes fainter than SNeI-a in spirals,
if one accepts as zero point M$_B=-19.53\pm 0.07$
(Tammann et al. 1996), the absolute magnitude at maximum of SNeI-a
in Spirals.

It is worthwhile noting in this respect that recent work by Mazzali et
al.  (in preparation) shows a good correlation between the velocity
widths of the nebular lines (at around 300 days after the maximum) 
and the rate of decline
(and therefore the absolute magnitude at maximum). The velocity widths
for SN 1992A would indicate that it is subluminous relative to other
normal type Ia supernovae.

The close similarity with the apparent magnitude at maximum exhibited
by SN 1980N and SN 1981D, two other SNeI-a in the Fornax Cluster,
could indicate that SN 1992A is a peculiar object only
if NGC 1316,
the parent galaxy of SN 1980N and 1981D, is placed on the far side of
the cluster at d$\gsim 20.5$ Mpc. In this case the absolute magnitudes
of SN 1980N and SN 1981D would be slightly brighter than M$_B\sim -19$,
in agreement, within the errors, with the value reported by Branch,
Romanishin and Baron (1996) for SNeI-a occurring in early type
galaxies.
Finally, we are not certain that all necessary corrections were made
to the TO magnitude of the GCLF to guaranty its reliability as
standard candle. Indeed, in section 3 we discussed systematic effects
which could conspire to make the TO magnitude of NGC 1380 fainter by
$\sim 0.3$ mags.

We have derived the absolute magnitude at maximum of a number of
historical SNeI-a which have occurred in early type galaxies, whose
distances are known through the GCLF. Together with high quality data,
collected from the literature, we have produced a linear fit to the
data points in the $\Delta m(15)$ {\sl vs.} $M_B$ plane. We find a
slope similar to that measured by Phillips (1993). The $\sim 0.4$ mag
difference in the intercept probably reflects the zero point difference
existing between the distances obtained via Cepheids/GCLF (adopted in
this paper) and Surface Brightness Fluctuation and Tully-Fisher
methods (adopted by Phillips 1993).  After excluding SN 1991bg from
the fit and adopting the slope derived by Hamuy et al. (1996) we have 
determined 
a new zero point for this relationship and H$_\circ=62\pm 6$ km s$^{-1}$
Mpc$^{-1}$. It is apparent that any correction for reddening would tend 
to increase the obtained value of H$_\circ$.
Our plot in Fig. 6 shows that: 

1) SNeI-a in Spirals are located in the
bright and slow part of the diagram whereas the SNe in early type
galaxies fall in the faint and fast part of the plot. Owing to the
intrinsic dispersion of the relationship, it is apparent that an
analysis based only on SNe discovered in Spirals or in early type
galaxies would hardly reveal any correlation between the rates of
decline and absolute magnitude at maximum, then indicating that the
dependence of the absolute magnitude at maximum on $\Delta m(15)$ and
on the {\sl Hubble type} of the parent galaxies are almost equivalent.
On the basis of data reported in Tab. 5 (but excluding 1991bg), the
unweighted mean of the absolute magnitude at maximum of SNe Ia in early
type galaxies is M$_B=-19.05\pm 0.16$. By comparing this figure with
M$_B=-19.53\pm 0.07$ (Tammann et al. 1996) for Ia in Spirals, we obtain
$\Delta M=0.48\pm 0.17$. 

2) Sub-luminous objects such as 1991bg would indicate that the
M$_B$ {\sl vs.} $\Delta m(15)$ relationship becomes nonlinear 
for high rate of decline (see also the case of SN 1992K pointed out by
Hamuy et al. 1995). 
 
3) We find a difference of $\Delta m=0.31 \pm 0.14$ mag between the
absolute magnitude at maximum of SNeI-a calibrated with GCLF and PNLF
(see Tab. 6 and Tab. 7). To check the significance of this result we
have compared the distance moduli of the ten galaxies for which the
distances have been determined via PNLF and GCLF (see Tab. 1 of Jacoby
(1997) and Whitmore 1997). By using the TO mag reported in our Tab. 3
we find a difference (m-M)$_{GC}$--(m-M)$_{PNe}=0.56 \pm 0.24$ (and
(m-M)$_{GC}$--(m-M)$_{PNe}= 0.61 \pm 0.19$ after excluding NGC 3379).  
This result may put doubt
on the consistency of the zero points of these distance
indicators. However, if we assume as calibrator of the absolute
magnitude at maximum of SNeI-a in early type galaxies the data
provided by PNLF ( Tab. 7), the price to pay to fit the AMMRD with
0.784 slope (Hamuy et al. 1996) is to make considerably fainter the
absolute magnitude at maximum of SNeI-a in spirals, close to M$_B\sim
-19$. We note that Kennicutt et al. (1998) have estimated
that metallicity effects on the distance scale derived with HST
observations of Cepheids could affect the distance moduli, to the
respective parent galaxies, only at $\sim 0.2$ mag level.

At least four critical observations could significantly improve our present
understanding of the problem:
 
a) determination of the distance of NGC 1316, parent galaxy of SN 1980N
and 1981D, through the use of the GCLF (the PNLF distance
already being available). This should also enable us
to clarify whether or not SN 1992A is a peculiar (sub-luminous)
object, despite its `protonormal' spectroscopic evolution.

b) determination of the distance of NGC 1380 via the 
PNLF to make a sensible comparison
with the distance obtained via GCLF in this paper. 

c) determination of the distance to NGC 4526, parent galaxy of SN 1994D,
through the use of the GCLF and PNLF. This SN may be a conspicuous exception
to the AMMRD relationship, unless we assume that NGC 4526 is
located on the near side of the Virgo cluster at $\sim 13$ Mpc (see
Tonry 1995).  However, this last possibility appears quite unlikely,
because it is known that early type galaxies are normally concentrated
towards the core of the clusters. As an alternative, NGC 4526 may
not belong to Virgo Cluster, rather being a foreground galaxy.

d) measurement of the distance, with Cepheids, to NGC 3627 and NGC 4527,
parent galaxies of SN 1989B and SN 1991T.  Indeed, these objects have been well
studied, close to maximum, in the past 
(Barbon et al.  1990, Wells et al. 1994, Phillips
et al. 1992, Ruiz-Lapuente et al. 1992, Filippenko et al. 1992b), and 
therefore,  once their position in the M$_B$ {\sl vs.} $\Delta m(15)$ plane is
firmly established, the large error still associated with the intercept of [1]
will be considerably reduced.

\subsection*{Acknowledgments}

The authors are strongly indebted to Tom Richtler for his generous
help in studying the Globular Cluster Luminosity Function of NGC
1380 and to an anonymous referee for comments which
helped to improve the manuscript.  Sergio Ortolani and Nino Panagia
provided useful comments on a preliminary version of this paper.
This research made use of the NASA/IPAC extragalactic database (NED)
which is operated by the Jet Propulsion Laboratory, Caltech, under
contract with the National Aeronautics and Space Administration.

\newpage


\begin{figure}
\psfig{figure=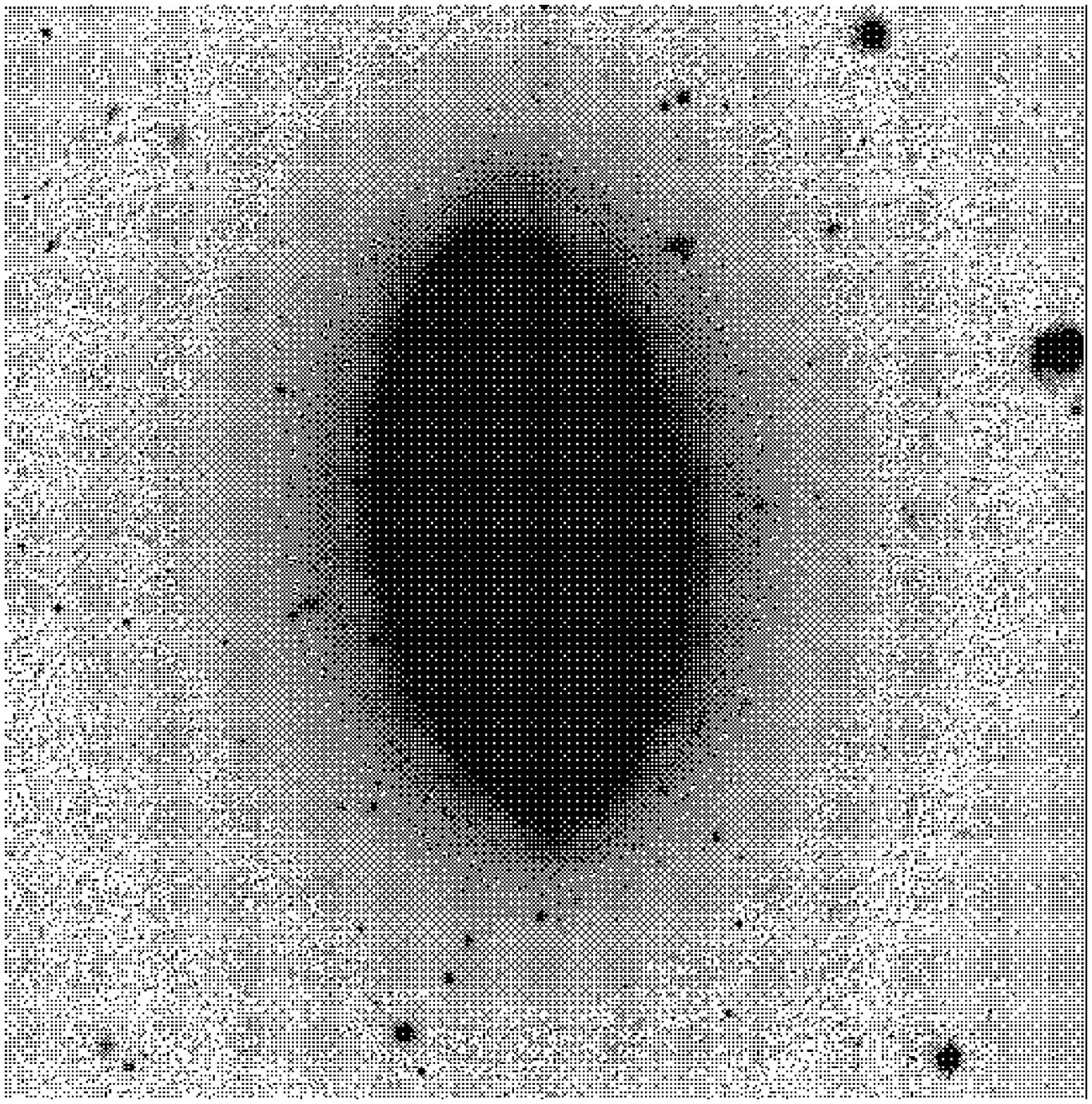,height=8cm,width=8cm
,bbllx=8mm,bblly=57mm,bburx=205mm,bbury=245mm}
\caption{The NGC 1380 galaxy (B 15000 s).}
\end{figure}

\begin{figure}
\psfig{figure=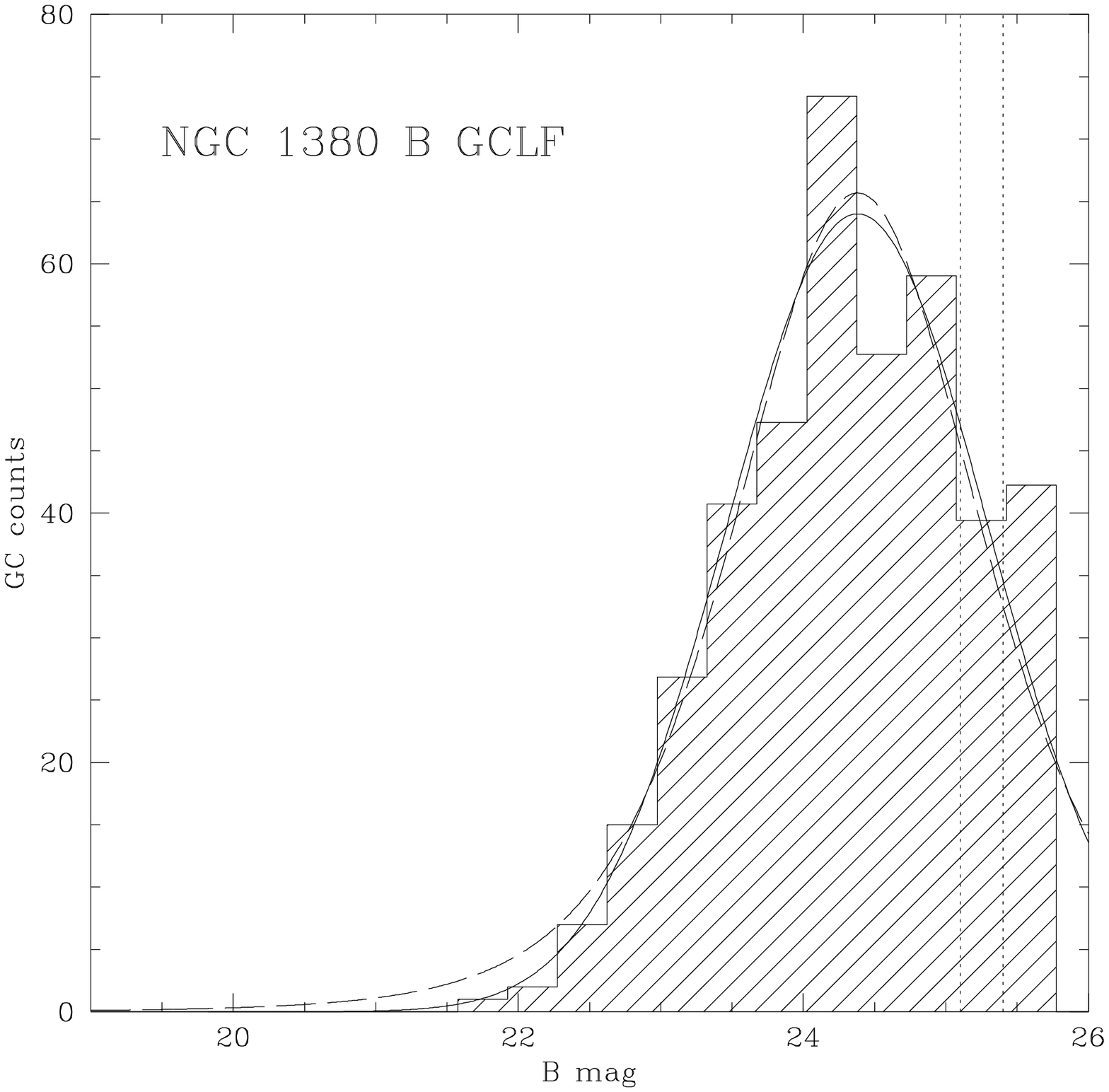,height=8cm,width=8cm
,bbllx=8mm,bblly=57mm,bburx=205mm,bbury=245mm}
\caption{ Globular cluster luminosity function in $B$. The solid curve
shows the best Gaussian fit, the dashed curve the best fit of a $t_5$
function. The dotted lines show the 60\% completeness limit in $B$
(left) and in $BVR$ (right).}
\end{figure}

\begin{figure}
\psfig{figure=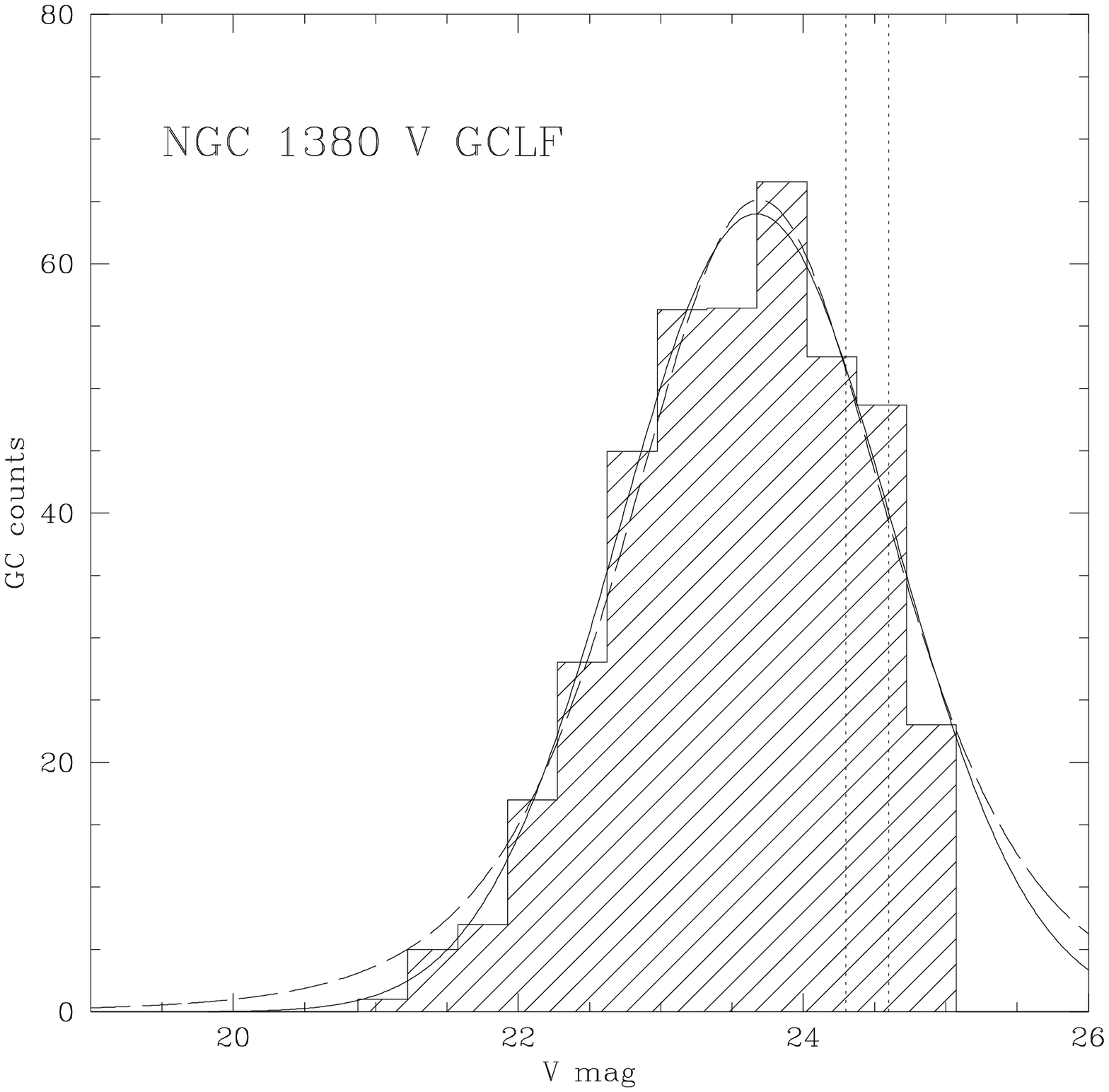,height=8cm,width=8cm
,bbllx=8mm,bblly=57mm,bburx=205mm,bbury=245mm}
\caption{ Globular cluster luminosity function in $V$. Symbols as for $B$.}
\end{figure}

\begin{figure}
\psfig{figure=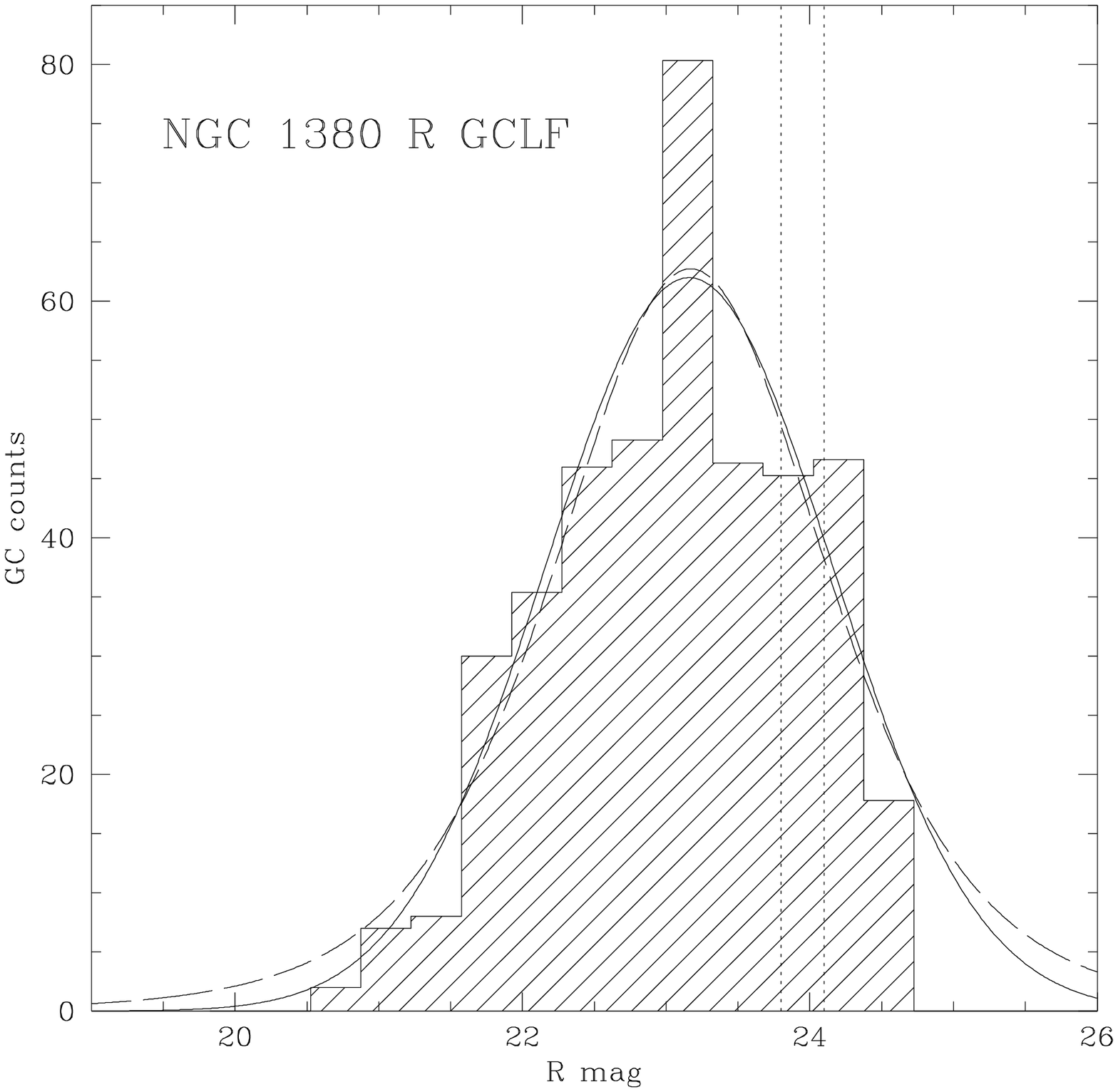,height=8cm,width=8cm
,bbllx=8mm,bblly=57mm,bburx=205mm,bbury=245mm}
\caption{ Globular cluster luminosity function in $R$. Symbols as for $B$.}
\end{figure}

\begin{figure*}
\psfig{figure=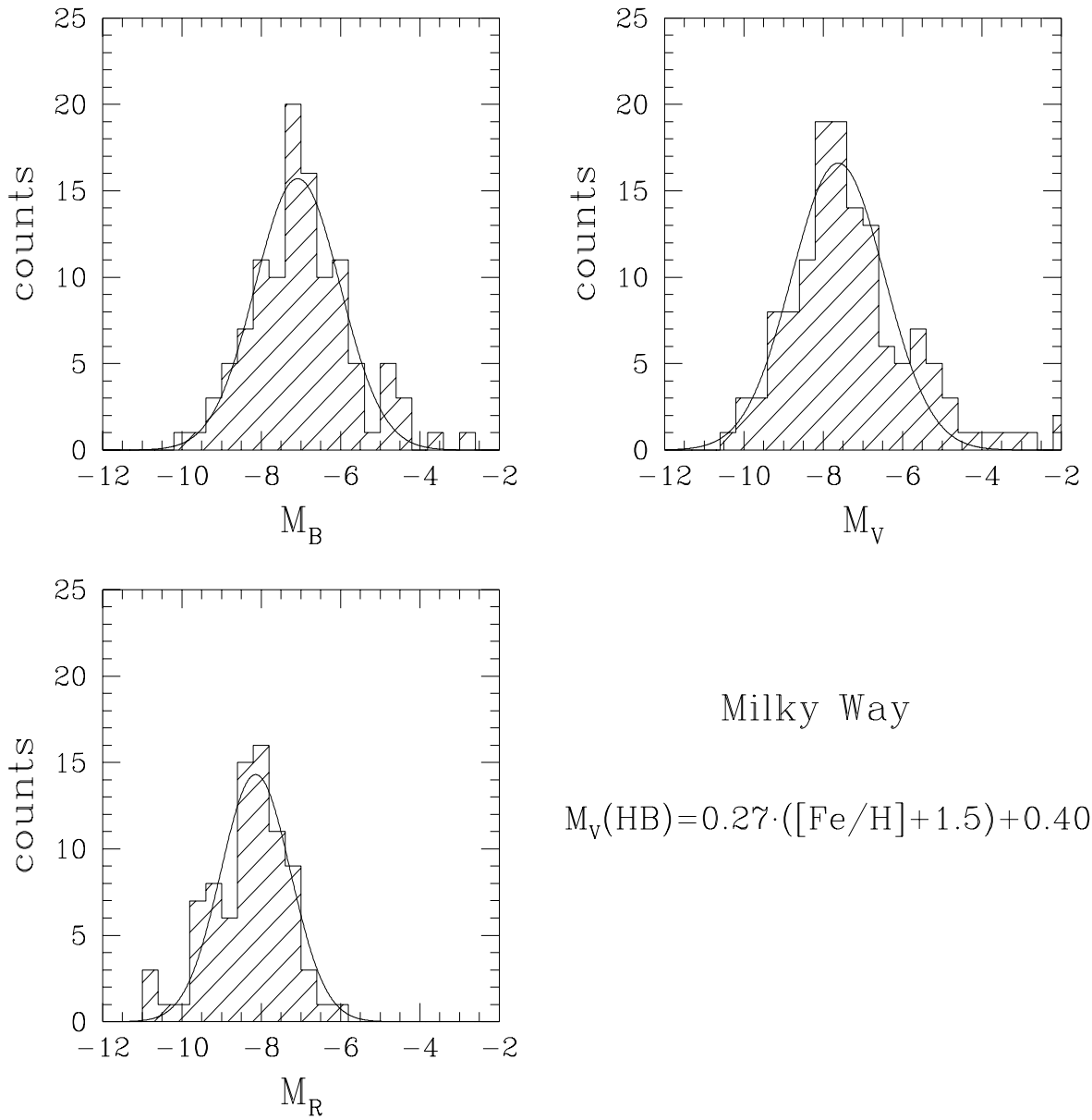,height=16cm,width=16cm
,bbllx=8mm,bblly=57mm,bburx=205mm,bbury=245mm}
\caption{ Globular cluster luminosity functions in $B,V,R$ of the Milky
Way as derived from the McMaster University globular cluster database.}
\end{figure*}

\begin{figure*}
\rotate[r]{
\psfig{figure=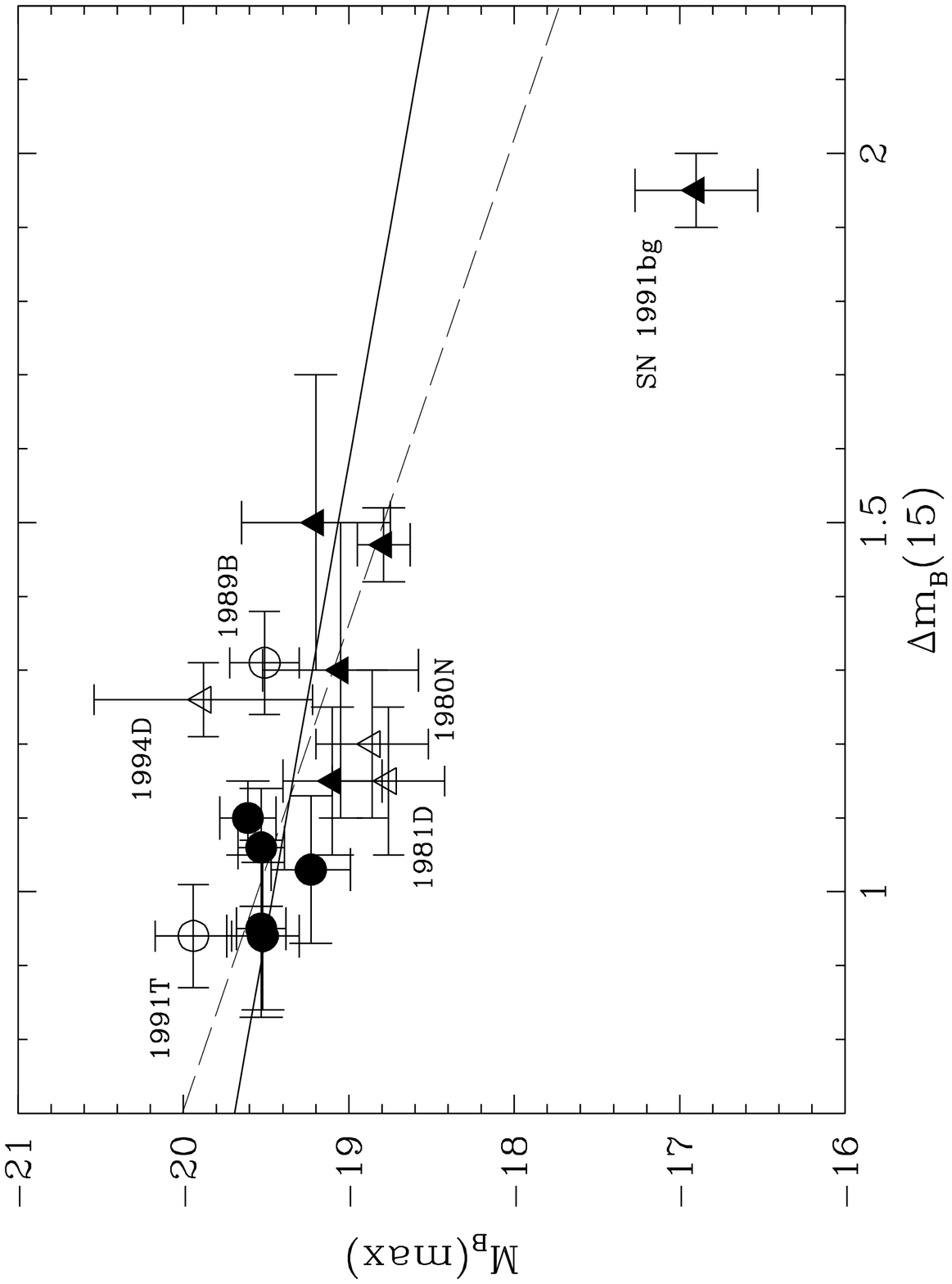,height=16cm,width=16cm
,bbllx=8mm,bblly=57mm,bburx=205mm,bbury=245mm}
}
\caption{The Absolute Magnitude at Maximum {\sl vs.} Rate of decline
relationship for SNeI-a. The filled symbols represent the objects
whose individual distances have been measured via Cepheids (circles)
or GCLF (triangles).  The light symbols represent SNe for which the
individual distances via Cepheids or GCLF are not available, therefore
the average distance of the respective cluster has been used.
These SNe have not been used to compute the best fit. The solid line
represents the best fit (without SN 1991bg) after assuming the Hamuy
et al.'s (1996) slope of 0.784. The dashed line is the best-fit
obtained by using a least squares fit which takes into account the
errors in both coordinates (slope 1.52).}
\end{figure*}

\end{document}